\documentclass{article}
\usepackage{amsmath}
\usepackage{amssymb}
\usepackage{cite}

\def\be{\begin{eqnarray}}
\def\ee{\end{eqnarray}}
\def\bes{\begin{eqnarray*}}
\def\ees{\end{eqnarray*}}
\def\ba{\begin{array}}
\def\ea{\end{array}}
\def\beenu{\begin{enumerate}}
\def\eenu{\end{enumerate}}
\def\bt{\begin{tabular}}
\def\et{\end{tabular}}
\def\beitem{\begin{itemize}}
\def\eitem{\end{itemize}}

\def\tr{\mbox{tr }}

\def\tr{{\mbox{tr}\,}}
\def\dim{{\mbox{dim}\,}}

\def\diff#1#2{{\frac{\partial #1}{\partial #2 }}}
\def\l|{\left| \vphantom{\bigl(}\right. }
\def\r|{\left. \vphantom{\bigl(} \right|}

\def\ti{\tilde \imath}
\def\tj{\tilde \jmath}
\def\ta{\tilde \alpha}

\newcommand{\Mz}{${\cal M}_z$}
\newcommand{\MI}{${\cal M}_I$}
\newcommand{\K}{ {\cal K} }
\newcommand{\C}{\mathbf{C}}
\newcommand{\Fsu}{$SU(N_f)\times SU(N_f) \times U(1)_B \times U(1)_R$}

\def\latin#1{{\it #1}}

\def\ie{\latin{i.e.}}

\def\apriori{\latin{a priori}}

\def\myskip{\vskip.1cm}

\newcommand{\drawsquare}[2]{\hbox{%
\rule{#2pt}{#1pt}\hskip-#2pt%  left vertical
\rule{#1pt}{#2pt}\hskip-#1pt%  lower horizontal
\rule[#1pt]{#1pt}{#2pt}}\rule[#1pt]{#2pt}{#2pt}\hskip-#2pt%  upper horizontal
\rule{#2pt}{#1pt}}% right vertical

\newcommand{\Ysymm}{\raisebox{-.5pt}{\drawsquare{6.5}{0.4}}\hskip-0.4pt%
        \raisebox{-.5pt}{\drawsquare{6.5}{0.4}}}%  symmetric second rank
\newcommand{\Yasymm}{\raisebox{-3.5pt}{\drawsquare{6.5}{0.4}}\hskip-6.9pt%
        \raisebox{3pt}{\drawsquare{6.5}{0.4}}}%  antisymmetric second rank

\begin{document}

%%%%%%%%%%%%%%%%%%%%%%%%%%%%%%%%%%%%%%%%%%%%%%%%%%%%%%%%%%%%%%%%%%%%%%%%%%%%%%%%
\begin{titlepage}

\begin{flushright}
Saclay T98/090 \\
hep-ph/9808345 \\
\end{flushright}

\vskip.5cm
\begin{center}
{\huge{\bf Anomaly Matching and Syzygies in $N$=1 Gauge Theories}}
\end{center}
\vskip1.5cm

\centerline{ Ph. {\sc Brax} , C. {\sc Grojean} 
{\it and} 
C.A. {\sc Savoy} }
\vskip 15pt
\centerline{CEA-SACLAY, Service de Physique Th\'eorique}
\centerline{F-91191 Gif-sur-Yvette Cedex, {\sc France}}
\vskip 2cm 

\begin{abstract}
We investigate the connection between the  moduli space of $N=1$
supersymmetric gauge theories and the set of polynomial gauge
invariants constrained by classical/quantum relations called {\it
syzygies}. We examine the  existence of a superpotential reproducing
these syzygies and the link with  the 't~Hooft anomaly matching between
the fundamental fields at high energy and the gauge invariant degrees
of freedom at low energy for the flavour symmetry group.  We show that
the anomaly matching is equivalent to the vanishing of the flavour
anomaly on the normal space to the manifold defined by the syzygies.
For normal spaces in a real representation of the flavour group we
strengthen the connection  between the 't~Hooft anomaly matching and
the existence of a superpotential by  constructing a flavour invariant
polynomial whose gradient vanishes at least on the solutions of the syzygies.
This corroborates a recent definition of confining theories.  We
illustrate our general result by considering two examples based on the
$SU(N_c)$ and $Spin(7)$ gauge theories.  We also examine  the role of
syzygies in the context of non-Abelian duality.  We emphasize the
relevance of non-perturbative effects in the dual magnetic theories in
proving the equivalence of the electric and magnetic syzygies.
\end{abstract}
\vfill

\vfill
\footnoterule
\noindent
e-mail addresses:
brax, grojean, savoy@spht.saclay.cea.fr

\end{titlepage} 

%%%%%%%%%%%%%%%%%%%%%%%%%%%%%%%%%%%%%%%%%%%%%%%%%%%%%%%%%%%%%%%%%%%%%%%%%%%%%%%%
\section{Introduction}
\label{sec:intro}

Recently there has been a tremendous increase of one's understanding of
supersymmetric field theories in four dimensions. Extended
supersymmetric theories with $N=2,4$ provide a wealth of information on
the non-perturbative nature of quantum field theories. The $N=1$
theories have also been thoroughly investigated. Nevertheless their
description is not as complete as their extended counterparts. This is
certainly due to the less restrictive mathematical framework of $N=1$
theories. Yet, a new insight on the infrared (IR) properties of
asymptotically free supersymmetric gauge theories has been provided by
the recent work \cite{Seiberg:1994bz, Seiberg:1995pq} on these
theories\footnote{ For a complete review on first duality examples
see \cite{Intriligator:1996au, Peskin:1997qi, Shifman:1995ua} and  many
other examples constructed later have been reviewed in
\cite{Brodie:1996xm}.}.  The most striking result on $N=1$ theories is
the existence of a new type of duality. This duality relates two
apparently different theories in the short distance regime that are
described by the same effective theory in the IR limit. In the same
vein the basic question concerning the  issue of colour confinement has
been tackled and clarified in these non-perturbative approaches for a
large class of supersymmetric theories. The dual descriptions and the
strong coupling effects have also received a new treatment and a
broader  understanding  via the study of the  D-brane dynamics
\cite{Elitzur:1997fh}.

The key ideas in these studies are the non-renormalization theorems and
the holomorphy of the superpotential on the one hand
\cite{Shifman:1986zi, Amati:1988ft, Shifman:1991dz, Seiberg:1993vc},
and the matching of the global symmetry anomalies as advocated by
't~Hooft \cite{'tHooft:1980xb} on the other hand.  Effective theories
have been written and argued to describe the IR behaviour of some gauge
theories in terms of gauge invariant composite chiral superfields
\cite{Amati:1988ft, Seiberg:1994bz, Seiberg:1995pq}.  The perturbative
quantum corrections only affect the K\"ahler metrics through
wave-function renormalization,  the non-perturbative quantum effects
are fixed to a great extent by the  holomorphy and by the global
symmetries of the theory.  The remaining ambiguities in the IR theory
are lifted and further support for their validity is obtained through
the use of several arguments: the decoupling of flavours (chiral
multiplets) establishing a descent link among a series of theories with
the same gauge group and different representations for the chiral
supermultiplets, the reduction of the gauge symmetry through the Higgs
mechanism.

The relevance of the analytic gauge invariants in the study of
supersymmetric theories has been noticed a long time ago.  Indeed, the
$F$-terms in the scalar potential are components of the gradient of the
superpotential, an invariant analytic function.  The $D$-terms are
Hermitean functions of the scalar fields, but a beautiful result in
algebraic geometry \cite{Kempf} provides the link with holomorphic invariants,
 at least for the analysis of the vacua 
\cite{Buccella:1982nx, Procesi:1985hr, Gatto:1985jz, Luty:1996sd}.
The degeneracy of the classical
solutions is described in terms of massless fields, the {\it moduli}. The
fundamental mathematical property of the (classical) moduli space
(modulo the group of gauge symmetry) is its isomorphism with an
algebraic variety of analytic gauge invariant polynomials in the
primordial fields, which we call the {\it chiral ring,} following
ref.\cite{Seiberg:1995pq}.  The chiral rings can be classified into two
categories, those which have algebraic constraints among the invariants
of the integrity basis, called {\it syzygies,} and those with a free
basis.  This classification is also related to the possible patterns of
gauge  and flavour symmetry breaking and to the structure of the
commutant, the largest symmetry group 
 commuting with the gauge group.  These results are summarized in the section 
\ref{sec:flat} of this paper.

A powerful and necessary criterion for the existence of an effective
theory describing the IR regime of an asymptotically free gauge theory
was stated by 't~Hooft :  there should be  matching of the (formal)
anomalies of the global symmetries calculated with either the UV or the
IR massless fermions. This has been extensively exploited to classify
confining theories (for a review see \cite{Csaki:1998th}).  In section
\ref{sec:anomalies}, we also stress that the 't~Hooft matching
conditions for the {\it unbroken global (flavour) symmetries} have to
be satisfied at the origin of the UV and IR field spaces. It is then
easy to select the unbroken flavour subgroups that are allowed at low
energies. The breaking has to be provided by an appropriate
superpotential which is restricted by the global symmetries, including
a $R$-symmetry.

One of the main aims  of this paper is the  analysis of the necessity
of a sufficient condition recently obtained \cite{Dotti:1997cr} from
the isomorphism between  the moduli space and the chiral ring
constrained by the syzygies.  With this completion the conjecture
reads:

\myskip 
{\it The 't~Hooft anomaly matching conditions are satisfied for supersymmetric gauge theories if and only if the syzygies of the chiral ring derive from a superpotential.}

\myskip
Rather than proving this statement we shall be interested in
making explicit the intimate link between a physical condition,
the t'Hooft anomaly matching, and the geometry of the space of
gauge invariants. 
We show that the anomaly matching is equivalent to the vanishing
of the flavour anomaly on  the normal space to the zeros of the syzygies.
In the case where  the normal space is 
real under the action of the residual flavour subgroup, we construct a flavour
group invariant polynomial in the gauge invariants whose gradient
vanishes at all the solutions of the syzygies. For the proof to be
complete it remains to be shown that all of the zeros of the gradient
of this superpotential are solutions of the syzygies.  Now, as
discussed in section \ref{sec:superpotential}, if a superpotential
exists that reproduces the syzygies, the normal space transforms as a
real representation of the flavour group, implying that the
correspondence between the anomaly matching condition and the existence
of superpotential occurs in  the widest  possible sense.

The global  symmetries of a supersymmetric gauge theory always possesses an Abelian
subfactor $U(1)_R$, the $R$ symmetry, which differs notably from
the rest of the flavour symmetries.
The role of the $R$ symmetry has been crucial in analyzing both
confining theories and the $N=1$ duality. The main difference
between the $R$ symmetry and the flavor symmetries stems from the
non-trivial action of the $R$ symmetry on the gauginos , the fermions in the gauge
vector multiplets, while   the flavour symmetries leave the
vector multiplets invariant.
Another essential property of the $R$ symmetry is the different
$R$ charges of the bosons and fermions in a chiral multiplet. The
boson have a $R$ charge which is shifted by one unit from the
charges of the fermions, \ie\ $R_F=R_B-1$.
This plays a noteworthy role in the analysis of the matching
conditions as the fermions are the fields involved in the anomaly
calculations.

At low energy we shall only consider the cases where the gauge
invariant polynomials in the matter fields are sufficient to
satisfy the matching conditions.   
In the rest of this paper we shall not be interested in the
explicit role that the invariants constructed using the field
strength chiral superfield $W_{\alpha}$ can have.

It is well-known that the $R$-symmetry only allows for a polynomial
superpotential for chiral supermultiplets in representations of the
gauge group with special values of Dynkin index, and the corresponding
theories have been extensively analysed \cite{Dotti:1998rv,
Dotti:1998wn, Csaki:1997zb, Grinstein:1998zv, Grinstein:1998bu}. Our
result corroborates the conjecture that the confining theories are
those that admit a polynomial superpotential \cite{Csaki:1997sm,
Csaki:1997zb}. These proofs are given in section
\ref{sec:superpotential}.

The second object of this paper is to present a consistency check on dual
theories: we verify that the solutions of the magnetic theories satisfy
the syzygies of the electric theory if and only if a non-perturbative
superpotential is added to the former.  This is done by considering the
theories with any number $N_f$ of chiral multiplets in the fundamental
representations of the gauge groups $SU(N_c)$ and $Spin(7)$. We give
the (unique) superpotentials for any $N_f$ which are consistent with
the flavour symmetries and with the successive decoupling of flavours.
Then we find  the gauge invariants and their syzygies and proceed
to test the electric syzygies in terms of the magnetic solutions. These
results are displayed in sections \ref{sec:examples} and \ref{sec:duality}.

Finally, in section \ref{sec:geometry}, the geometry of the moduli
space of the $SU(N)$ theories with a large number of flavours is
determined in an attempt to have access to the (classical) K\"ahler
potential of the dual theory.  This is done by finding the non-compact
invariance group of the equations of motion. It is shown that the dense
part of the moduli space consists of two conjugated orbits of this
non-compact group whose closure are the singular orbits. This provides
a coset parameterisation of the solutions which, however, is not
(explicitly) holomorphic and breaks the K\"ahler invariance.  Therefore
it is not possible to obtain the induced K\"ahler potential for the
dual theory in a straightforward way. But some issues suggested by
these results are under investigation.

%%%%%%%%%%%%%%%%%%%%%%%%%%%%%%%%%%%%%%%%%%%%%%%%%%%%%%%%%%%%%%%%%%%%%%%%%%%%%%%
\section{Flat Directions, Analytic Invariants and Syzygies}
\label{sec:flat}

In this section we recall some important mathematical results that
allow  a general analysis of supersymmetric gauge theories.  In global
supersymmetric theories the scalar potential is a sum of $F$-- and
$D$--terms~: the $F$--terms are the norm of the gradient $W_i$ of a
holomorphic function, the superpotential $W(z^i)$. For the $D$--terms,
the relation with holomorphy is more subtle and appears when we are
interested in supersymmetric vacua.  In a supersymmetric theory with
scalar fields $z^i$, K\"ahler  potential $\K (z,z^\star)$ and gauge
group $G$, the $D$--terms are defined as
\be
D^A = \diff{\K}{z^i} \left( T^A z \right)^i
\ee
where $T^A$ $(A=1,2, \ldots ,\dim G)$ are the Hermitian generators of $G$
in the representation of the scalar fields.

A sufficient and necessary condition for the vanishing of the $D$--terms is
\cite{Buccella:1982nx, Procesi:1985hr, Gatto:1985jz}:

\myskip 
\noindent {\it For any holomorphic gauge-invariant polynomial $I (z^i)$
in the scalar fields, a solution $\xi$ of the equation
\be
\diff{I}{z^i}_{|z=\xi} = C \diff{\K}{z^i}_{|z=\xi}
\label{invariant}
\ee
where $C$ is a complex constant, is a solution of $D^A=0$ (of course
the whole $G$--orbit associated to $\xi$ is a solution). Conversely to
any solution of $D^A=0$ one can associate a holomorphic gauge invariant
satisfying (\ref{invariant}).}
\myskip

\noindent The proof of this results is obtained by studying the closed
orbits of the complexified $G^c$ of the gauge group $G$ and the ring of
$G^c$--invariant analytic polynomials.  This ring is finitely
generated~: one can find an integrity basis \ie\ a set of
$G$--invariant holomorphic homogeneous polynomials $\left\{ I^a (z)
\right\}_{a=1\cdots d}$ such that every $G^c$-invariant polynomial in
$z$ can be written as a polynomial in the $I^a (z)$.

 The elements of  an integrity basis are not always algebraically
independent.  In general, there exist algebraic relations (called {\it
syzygies}) satisfied by the fundamental invariants \footnote{ A
trivial example is provided by the $SU(N_c)$ gauge theory with $N_c$
fields in the fundamental and antifundamental representations; the
fundamental invariants are the mesons $M=z{\tilde z}$ and the baryons
$B=\det z$ and  ${\tilde B}= \det {\tilde z}$; classically, they are
constrained by $\det M- B {\tilde B}=0$. See below for further
examples.} :
\be
{\cal S}^{\alpha} \left( I^a (z) \right) = 0 \ \ \ \ \alpha=1,\ldots ,
N \label{syzygies} 
\ee 
To each $G^c$--orbit corresponds a vector in $\C^d$ made out of the
values taken by the invariants $\left\{ I^a (z) \right\}$ along this
orbit.  Conversely, it can be shown that to each vector in $\C^d$
satisfying the syzygies\ (\ref{syzygies}) is associated a unique closed
$G^c$-orbit.  In that sense the algebraic manifold defined by the
syzygies\ is identified with the set of closed $G^c$--orbits.  Notice
that the origin $\left\{ I^a = 0 \right\}$ is associated with the
unique closed $G^c$--orbit of $z^i = 0$. 

The existence of the syzygies can be related to the index of the matter
field representation,  $\mu=\sum_i\mu_{i},$ where $\mu_{i}$ is the
index of the irreducible representation $R_i$ defined by
$\tr({T^a(R_i)T^b(R_i)})=\mu_{i}\delta^{ab}.$ For low indices
$\mu<\mu_{adj},$ where  $\mu_{adj}$ is the index of the adjoint
representation, the gauge invariant rings have been classified
\cite{Schwarz:1978gw, Adamovich:1983, Dotti:1998wn, Dotti:1998rv} and
it has been shown that there are no syzygies.  For indices larger than
the index of the adjoint representation the generic situation is that
there are syzygies, with a few exceptional cases with no syzygies.

Equations (\ref{invariant}) can be seen  as a condition for the points
of a closed $G^c$-orbit to extremize the K\"ahler potential, the
constant $C^{-1}$ being a Lagrange multiplier.  A result of geometric
invariant theory \cite{Kempf} states that the points extremizing the
K\"ahler\ potential on a $G^c$-orbit form a unique $G$-orbit and are
solutions of $\left\{ D^A = 0 \right\}$.  Identifying   the points on a
same $G$-orbit,  there is a one-to-one correspondence between any two
of the following sets~:

\myskip
{\it
(i) the algebraic manifold defined by the syzygies\ (\ref{syzygies});

(ii) the closed $G^c$-orbits;

(iii) the solutions of (\ref{invariant}) modulo gauge transformations;

(iv) the solutions of $D^A=0$ modulo gauge transformations.}
\myskip

Therefore, there is a homeomophism between the set of the scalar field
space defined by $D^A=0$, \ie\ the moduli space, \Mz , and the
algebraic variety defined by  the fundamental invariants $I^a$
constrained by ${\cal S}^{\alpha} \left( I^a (z) \right) = 0$, \MI :
\be
{\cal M}_I \sim {\cal M}_z .
\label{iso}
\ee 
\noindent This homeomorphism is instrumental in studying the low
energy physics of supersymmetric gauge theories.

There is a natural stratification of \Mz according to the unbroken
gauge symmetry, {\it i.e.,} the little group $G_z \subset G$ at each
point $z$. A {\it stratum} $(G_z)$ is the set of all points which have
the same invariance $gG_zg^{-1}$, where g is any element of $G$. The
generic or principal stratum has the smallest residual symmetry $G_z$
while the more singular ones correspond to larger numbers of unbroken
gauge symmetries.  For $\mu<\mu_{adj }$ the gauge group is generically
broken to a non-void residual gauge group, while for $\mu > \mu_{adj}$
the gauge group is totally broken in the generic stratum.  Because of
the isomorphism (\ref{iso}), there is a corresponding stratification of
the algebraic manifold defined by the syzygies, \MI .

A powerful property for the study of supersymmetric gauge theories is
the following theorem \cite{Procesi:1985hr, Abud:1981}~: 

\myskip
{\it The tangent space to the moduli space  at a point $\xi$ in a 
stratum is analytically isomorphic to the tangent space at the point
$I^a(\xi)$ to the corresponding stratum of the manifold of gauge
invariant polynomials.} 
\myskip

\noindent In particular, it has an important consequence on
 anomalies of the flavour group in both theories
and the 't~Hooft matching conditions.

%%%%%%%%%%%%%%%%%%%%%%%%%%%%%%%%%%%%%%%%%%%%%%%%%%%%%%%%%%%%%%%%%%%%%%%%%%%%%%
\section {The 't~Hooft Anomaly Matching Conditions}
\label{sec:anomalies}

Let us recall that the matching conditions concern the global
(unbroken) anomalous symmetries $H$ of a given gauge theory. It
specifies that the anomalies $\tr H^3$ and $\tr H$ are equal when
evaluated with the high energy content of the gauge theory, \ie\ the
matter fields $z^i$ in diverse representations of the gauge and flavour
groups, and with the low energy degrees of freedom of the theory.  The
(perturbative) degeneracy of the supersymmetric vacua  requires a more
careful analysis. In a supersymmetric theory with gauge symmetry group
$G$ and superpotential $W(z^i)$ with flavour symmetry $H$, at any
vacuum $z$ such that $W_i(z)=0$ and $D^A(z, z^{*}) =0$, the mass
matrices for the chiral multiplets are $W_{ij}(z)$ and $D^A_i(z,
z^{*})$ . If the residual global (flavour) group at $z$ is $H_z \subset
H$, the mass matrix is $H_z$--invariant and the massive states are in a
real (vector-like) representation of $H_z$. Therefore the anomalies
$\tr H_z^3$ and $\tr H_z$ can be calculated at $z$ where the massive
states are excluded or at the origin $z=0$ where all states are
massless. In particular the 't~Hooft conditions can be checked at the
origin of the UV theory on the one hand, and at the origin of the IR
theory on the other hand, for each relevant flavour subgroup $H_z
\subset H$.

The role of the superpotentials (both in the UV and the IR theories) is
then clear:  {\it they cannot change the anomaly matching, but they
could restrict the vacua, hence the flavour invariance of the theories
to a subgroup whose anomalies match,} provided that such
superpotentials that are consistent with the symmetries of the theory
exist. This argument corresponds to examining the anomaly matching at
the origin of the moduli space. Given the UV fields, $z^i$, and the IR
composites, $I^a$, the flavour symmetries consistent with the 't~Hooft
conditions are easily checked by considering the chains of subgroups
$H_z$ and calculating their anomalies at the origin.

Since the inclusion of a superpotential is easily taken into account,
we assume for simplicity that the UV theory has no superpotential.  As
seen in the previous section the moduli space is homeomorphic to the
variety of gauge invariant polynomials. These gauge invariants are
chiral superfields which are good candidates for a low energy
description of the gauge theory. The direct relationship between the
moduli space and gauge invariants entails a close relationship between
the IR and UV anomalies.  

Let us decompose \Mz \ and \MI \ into strata and then apply the
theorem quoted in the previous section.  The relevant point of this
theorem is the analyticity of the mapping. It implies that chiral
superfields in the tangent space of the $D^A=0$ set at $z$ are mapped
to chiral superfields in the tangent space of the manifold of gauge
invariants at $I^a(z)$. The tangent space within a given stratum
corresponds to  singlet fields under the residual gauge group. They
fall into representations of the residual global symmetry $H_z$ and are
mapped into corresponding representations in the tangent space to the
variety of gauge invariants.  An immediate consequence is {\it the
equality between the contributions to the anomalies from the fields
associated to both tangent spaces.} Our results in the next section 
rely on this powerful property.

We first discuss an example before embarking upon the general cases.
For simplicity we describe the case of the $SU(N_c)$ gauge theory with
$N_f<N_c$ flavours of quarks and antiquarks \cite{Affleck:1984mk}.
This theory possesses two global axial symmetries that we shall denote
by $SU(N_f)_A\times U(1)_R$. The charges of both the quarks and the
antiquarks are  $(N_f,1-N_c/N_f)$ . The chiral ring is generated by the
(meson) invariants $M_{i\tj}$ transforming under the axial flavour
group as $(\Ysymm + \Yasymm, 2(1-N_c/N_f))$. The anomalies of the
various possible $SU(N_f)_A\times U(1)_R$ subgroups calculated (at the
origins) for the quark and antiquarks and for the mesons do not match,
so that the low energy theory cannot preserve any flavour symmetry.  At
the generic stratum of the  $D^A=0$ manifold, the gauge group is broken
to $SU(N_c-N_f)$, and there is no residual global symmetries. At the
singular strata, the gauge group is broken to  $SU(N_c-r)$, $r<N_f,$
and the axial flavour group is broken to $U(N_f-r)$. The tangent spaces
to the strata are, respectively, all the quarks but $(N_f-r)$ families
in the fundamental representations of $SU(N_c-r)$, and all the mesons
$M_{i\tj}$ except those for which $i > r$ and $j>r$.  The global
anomalies of  $U(N_f-r)$ are easily checked to coincide. However, by
taking into account the other mesonic fields, which are not in the
tangent space this matching is destroyed. Therefore one has to look
for a theory where besides the  $SU(N_c-r)$ gauge theory with $(N_f-r)$
flavours, the IR physics is defined in terms of this restricted set of
mesons.

The only non-perturbative superpotential consistent with the flavour
symmetries is \cite{Affleck:1984mk}
\be
 W=(N_c-N_f)
\left( \frac {\det M_D}{\Lambda^{3N_c-N_f}} \right)^{1/(N_f-N_c)}
 \label{affleck}
\ee
implying ${\rm det}M \ne 0$ and restricting $M$ to the generic
stratum.  The singular strata are excluded as they must be because of
the anomaly mismatch. Actually, the only way to introduce a
supersymmetric theory with the mesonic fields restricted to the coset
$U(N_f)/U(N_f-r)$ would be through a non-linear realization of the
global symmetries, namely, through a non-trivial K\"ahler potential. Of
course, this is a difficult task in the absence of important
constraints such as holomorphy and perturbative non-renormalization
which allow for the determination of the superpotential. Anyway, as
such, the superpotential (\ref{affleck}) destabilizes the theory which
has no ground state.

On more general grounds, at  a non-trivial vacuum, the anomalies
receive two contributions, one from the charged fields under the
residual gauge group, $G_z$, and another one from the singlet fields.
Neither all singlet fields nor all chiral invariants belong to the
lesser dimensional tangent spaces to the singular strata which the
theorems of the previous section apply to, yet they may contribute to
the anomalies which may not match as in the previous example.  As
already stressed, the anomaly matching for the corresponding flavour
subgroups are easily checked at the origins.

Although the flavour symmetries that constrain the non-perturbatively
generated potential are model dependent, a general structure can be
already established from the general form of the non-anomalous
$R$-symmetry.  The charges of the matter fields $z_i$ under the
$U(1)_R$ can be chosen as
\be
R_i=1-\frac {\mu_{adj}}{n\mu_i} \label{Rcharge}
\ee
where $n$ is the number of matter field representations. The
charges of the fermions in the same chiral multiplets are $R_i-1=-\frac{\mu_{adj}}{n\mu_i}$.
The low energy superpotential is a combination of terms having
$R=2$ which can be expressed in terms of the matter field
content of the gauge invariants as \cite{Dotti:1998rv, Dotti:1998wn, Csaki:1997zb}
\be
W\sim (\prod_i z_i^{\mu_i})^{2/ ( \mu - \mu_{adj} )} \label{Wgeneral}
\ee
The special cases where $W=0$ have been classified and studied in Refs. 
\cite{Dotti:1998wn, Csaki:1997zb}.
When $\mu < \mu_{adj}$ the superpotential has a runaway behaviour
with a minimum at infinity. In a theory with confinement, this general
form has to be reproduced in terms of various combinations of the
chiral invariants, and there can be several. Its determination then
relies on different arguments involving decoupling and higgsing
\cite{Seiberg:1994bz, Seiberg:1995pq, Aharony:1995ne, Csaki:1997zb}. 
We give some examples of (\ref{Wgeneral}) in section \ref{sec:examples}

{}For low indices $\mu<\mu_{adj}$ the gauge invariant rings have been
classified \cite{Schwarz:1978gw, Adamovich:1983} and it has been shown
that there are no syzygies.  There were several prior examples of
confining theories of this kind in the literature \cite{Csaki:1998th,
Dotti:1997cr, Dotti:1998rv, Dotti:1998wn}.  The case study of all these
theories has not been carried out.  Nevertheless the studied cases
reveal the pattern emerging from the 't~Hooft matching, either the
anomalies do not match and there is a superpotential, or the anomalies
do match and there is a branch of the low energy theory with a
vanishing superpotential.  We refer to these detailed studies   and
concentrate on the $\mu \ge \mu_{adj}$ in the following.

%%%%%%%%%%%%%%%%%%%%%%%%%%%%%%%%%%%%%%%%%%%%%%%%%%%%%%%%%%%%%%%%%%%%%%%%%%%%%%%%
\section{Anomaly Matching and Syzygies}
\label{sec:superpotential}

The gauge theories with $\mu \ge \mu_{adj}$ have syzygies generically.
There are a few classified cases where the basic invariants are
not constrained. Barring these  examples which have been
extensively studied \cite{Dotti:1997cr, Dotti:1998wn, Dotti:1998rv, Grinstein:1998zv, Grinstein:1998bu} in the literature, 
in the following we shall be 
interested exclusively in theories
with non-void syzygies, \ie\ there exists relations between the
basic invariants of the theory.   
We shall now examine the connection between the anomaly matching and
the existence of a superpotential which reproduces the syzygies.

Let us recall that the moduli space ${\cal M}_z$ is homeomorphic to the
manifold ${\cal M}_I$ defined in terms of the composite fields $I^a$ by
the syzygies (\ref{syzygies}). The gauge group is generically broken,
\ie\ the gauge group is completely broken on the generic stratum, a
dense open set. In this section we concentrate on the principal
stratum. {}For the singular ones, where there remains a residual gauge
group, the analysis is more involved as already discussed in the
previous section.  Since the gauge group is completely broken, the
anomaly calculated from the chiral fields in the tangent space to the
moduli space at a point $z$ of the generic stratum is the same as the
anomaly (formally) calculated from the gauge invariants in the tangent
space to the corresponding point of the chiral ring with the syzygies
(\ref{syzygies}). Then~\cite{Dotti:1997cr}\footnote{See the remark at
the end of this section concerning the case $\mu=\mu_{adj}$ and
the quantum modifications.}:

\myskip
{\it The 't~Hooft anomaly matching conditions are
satisfied for supersymmetric gauge theories 
if  the syzygies ${\cal S}^a(I^a)=0$ of the gauge invariant ring
$ {\C [I^a]}/\{S^a=0\}$ derive from a superpotential $W(I)$ \ie\
${\cal S}^a=\partial W/ \partial I^a =0$.}   
\myskip

Let us show this result.  Assume that there exists a
superpotential $W(I^a)$ such that the syzygies are the $F$--terms 
\be
{\cal S}^{a}=\frac {\partial W}{\partial I^{a}} 
\label{sufnec}
\ee
The tangent space to the variety  defined by the sygyzies is given  by
the zero eigenvectors of the matrix 
\be
{\cal S}^b_a = \diff{{\cal S}^b}{I^a} (I^a) = 
\frac{\partial^2 W}{\partial I^a \partial I^b},
\label{MI}
\ee
\noindent namely, by the massless composite superfields.  The anomalies
in the IR supersymmetric theory  defined by the superpotential $W(I^a)$
are the same as the formal one defined within the tangent space to
${\cal S}^{\alpha}= 0$ .  As already stated, these coincide with the
anomalies calculated within the tangent space to the set of solutions
of  $D^A =0$, defined by the zero eigenstates of the matrix $D^A_i$,
namely, with the massless states of the UV supersymmetric theory. Of
course, the mass matrices are invariant under the residual global
(flavour) group  $H_z \subset  H$,  and the massive states are in a
real (vector-like) representation of $H_z$.  These anomalies of $H_z$
coincide also when calculated at the origins, $z^i=0$ and $I^a=0$,
respectively. Therefore the 't~Hooft matching conditions are
satisfied.

Let us examine the statement:  

\myskip
{\it  The 't~Hooft anomaly matching
implies the existence of a flavour invariant polynomial whose gradient
vanishes on the zeros of the syzygies. } 

\myskip
\noindent The 't~Hooft conditions imply the matching of the anomalies
of $H_z$ calculated in the fundamental theory with all the fields $z^i$
and in the low energy theory, with all the composite fields $I^a$.  The
former also match with the anomalies calculated with the zero
eigenstates of the mass matrix $D^A_i$, namely on ${\cal{M}}_z$, which,
in turn, match with the anomalies calculated with the zero eigenstates
of ${\cal S}^\alpha_a (I)$, namely on ${\cal{M}}_I$.  Therefore the
$N\times d$ matrix ${\cal S}^\alpha_a$ ($N<d$) projects onto the normal
space to ${\cal{M}}_I$ which transforms as an anomaly free
representation of $H_z$. There are two distinct possibilities
concerning the representations of $H_z$  forming the normal
space. When the normal space is an anomaly free chiral
representation of the little group there cannot exist a
superpotential reproducing the syzygies. This follows from our
earlier analysis showing that the existence of a superpotential
implies that the normal space is a real representation of the
little group.  In the following we only focus on the case where
the normal space is a real representation of $H_z$ and  show how this allows to construct a
$H$-invariant polynomial whose gradient is zero when the
syzygies are satisfied. It is important to notice that the
 restoration of the $H$-invariance of the constructed polynomial concerns the flavour
symmetries acting on the fermions associated to the invariants.
This implies the invariance of the polynomial under the $(R-1)$
Abelian symmetry.  

The transformations $\{ h\}$ of $H$ define the orbit $\{ hI \}$ of each
point $I \in \mathbb{C} [ I^a ]$. The covariance of the syzygies defines
a representation $\{ {\hat h} \}$ such that
\begin{equation}
{\cal S}^\alpha (hI) = {\hat h}^\alpha{}_{\beta}\, {\cal S}^\beta (I)
.
\end{equation}

At any point $I_0 \in {\cal M}_I$  the gradients ${\cal S}^\alpha_a
(I_0)$ are invariant under the action of the little group of $I_0$,
$H_{I_0}$.  It is then possible to construct a basis of the syzygies
and a basis of the invariants associated to the point $I_0$, with  the
indices $\alpha$ and $a$ decomposed into $\alpha=\{i,m\}$ and
$a=\{i,r\}$ such that:
\begin{equation}
\frac{\partial {\cal S}^\alpha}{\partial I^a} (I_0)
=
\left(
\begin{array}{cc}
D & 0 \\
0 & 0
\end{array}
\right)
\end{equation}
where $D$ is a non-singular diagonal matrix. The set $\{I^i\}$ spans ${\cal N}_{I_0}$, the normal space to
${\cal M}_{I}$ at the point $I_0$, and $\{I^r\}$, the tangent space.
Under the flavour group $H,$ 
\begin{equation}
I_0 \in {\cal{M}}_I \to hI_0 \in {\cal{M}}_I, \ \ \ 
H_{I_0} \to h H_{I_0} h^{-1}, \ \ \ 
{\cal N}_{I_0} \to h {\cal N}_{I_0} .
\end{equation}

The reality of ${\cal N}_{I_0}$ under $H_{I_0}$ implies the existence
of an unitary matrix $\eta$ such that for any element $h_{\cal N}$ of
the representation of $H_{I_0}$ on ${\cal N}_{I_0}$: $h^*_{\cal N} =
\eta h_{\cal N} \eta^{-1}$.  Now, if $I_0$ is the projection of $I$ on
${\cal M}_I$ (the projection is always defined at least in a
neighborhood of ${\cal M}_I$), we define $W(I)$ as the  integral along
the normal direction $(I-I_0)$~:
\begin{equation}
W(I)= \int_{0}^1
dt \,  
 {\cal S}^{i_1} (I_0 + t(I-I_0))\, (I-I_0)^{i_2}\, 
\eta_{i_1i_2} 
\label{superpote}
\end{equation}
where the basis associated to $I_0$ has been chosen.  By construction,
this function is $H_{I_0}$-invariant. To see its $H$-invariance, we
have to relate the basis at different points.  Consider any element $h$
of $H$. The unitarity of $H$ ensures that $hI_0$ is the projection of
$hI$ (note that the covariance of the syzygies guarantees that $hI_0$
belongs to ${\cal M}_I$).  It is easy to verify that ${\hat h}^{-1}
{\cal S}$ and $h^{-1}I$ are the bases associated to $hI_0$.  The
coordinates of the point $hI$ in the new basis are the same as those of $I$
in the old basis.  Moreover $\eta$ is independent of the basis.  Thus
\begin{equation}
\begin{split}
W(hI)\ & = \ \int_{0}^1
dt \,  
 ({\hat h}^{-1})^{i_1}{}_{i_2}
{\cal S}^{i_2} (h(I_0 + t(I-I_0)))\, (I-I_0)^{i_3}\, 
\eta_{i_1 i_3} 
\\
& = W(I)
\end{split}
\end{equation}
which proves the invariance of $W$ under the action of flavour group
$H$.  Note that the reality of ${\cal N}_{I_0}$, which we assume to
follow from the 't~Hooft matching condition, is necessary in our
construction of $W$; the invariance under $H$ then follows from
covariance. 
The $H$-invariance of the
polynomial $W$  concerns the flavour symmetries acting on the
fermions associated to the invariants. When $\mu\ne\mu_{adj}$ the $R$
symmetry is broken at low energy.  The
restored symmetry acting on $W$ is the $(R-1)$ symmetry  of the fermions.
Assuming that the directions $(I-I_0)^i$ have a well-defined $R$
charge, this leads to the invariance condition
$(R_S-1)+(R_I-1)=0$. The $R$ charge of the
polynomial $W$  is explicitly given by $R_W\equiv R_S+R_I=2$.
The $R$ symmetry of the polynomial $W$ is then two as required
for a superpotential.    

The gradients of $W(I)$  along the tangent directions to ${\cal M}_I ,$
\be
\partial_r W(I)=\int_0^1 dt 
{\cal S}^{i_1}_r(I_0+t(I-I_0))(I-I_0)^{i_2}\eta_{i_1i_2}
\label{grad1}
\ee
obviously vanishes when $I=I_0$.
Along the normal directions the gradient becomes
\be
\partial_i W(I)= {\cal S}_i(I)+ 
\int_0^1 dt 
({\cal S}_{i_1 i}(I_0+t(I-I_0))-{\cal S}_{ii_1}(I_0+t(I-I_0)))(I-I_0)^{i_1}
\label {grad2}
\ee
where the indices are lowered using $\eta$.  This vanishes for
$I=I_0$. This guarantees that the polynomial $W(I)$ has a vanishing
gradient on the zeros of the syzygies.  The previous formulae imply
that the gradient of $W$ is equal to the set of syzygies when the
matrix ${\cal S}^{\alpha}_a$ is symmetric. This requires a connection
between the $H$-representations of the syzygies and of the invariants
which seems difficult to establish in general.

Up to now our discussion has been restricted to the normal space
to one particular point  $I_0$ on ${\cal M}_I$. We have
constructed explicitly  a polynomial $W$ on the normal space at $I_0$ which
has a dependence on the  point $I_0$.
We shall now extend our analysis to non-singular strata 
 on  the  non-singular manifold $\hat {\cal M}_I$ of ${\cal M}_I$. By a standard result of differential
geometry \cite{bott} we can choose a tubular neighbourhood of
this manifold $\hat{\cal M}_I$ which is diffeomorphic to the normal
bundle ${\cal N}$. This is
an open neighbourhood of the manifold $\hat{\cal M}_I$. 
Locally in an open neighbourhood  the
coordinates describing the tubular neighbourhood are $(I_0,E)$
where $I_0\in \hat {\cal M}_I$ and the vector $E\in {\cal N}_{I_0}$ in the
normal space at $I_0$. This implies that every point in the tubular
neighbourhood is represented by the vector $I=I_0+E$.

 We can
extend the definition of $W$ to the stratum of $I_0$ on the
manifold $\hat{\cal
M}_I$. We choose $I_0$ such that its stratum possesses  a  little group
$H_{I_0}\ne 0$ with the smallest dimension.     
Let us now introduce the slice $\Sigma_{I_0}$ passing through
$I_0$ in the stratum of $I_0$. It is such that it parametrizes
the  orbits and is in a sense normal to them. In particular the
tangent space to the slice $\Sigma_{I_0}$ at $I_0$ comprises only
singlets of the little group $H_{I_0}$. Consider another point
$I'_0\in \Sigma_{I_0}\cap \hat{\cal M}_I$ then the little groups
$H_{I_0}\equiv H_{I'_0}$ and one can easily verify that the
normal spaces coincide.

Choose a point $I'$ in the tubular neighbourhood such that
$I'-I'_0=I-I_0$ identified with $E$ in the normal space. By the
line  integral similar to  \eqref{superpote} starting from $I'_0$  one can define a polynomial
$W'(I')$ with an explicit dependence on $I'_0$. One can also
extend the polynomial  $W(I)$ defined in the normal space at
$I_0$  to $W(I')$ using the Taylor polynomial 
in  $\delta I_0=I'_0-I_0$. A  calculation shows that the
two polynomials coincide
\be
W'(I')\equiv W(I')
\ee
This allows to extend the definition of the polynomial $W$ to the
whole tubular neighbourhood of the slice $\Sigma_{I_0}$ and
therefore by the $H$-invariance to the tubular neighbourhood to
the stratum of $I_0$ in $\hat{\cal M}_I$ .
We have then defined a polynomial $W$ on each of the tubular
neighbourhoods of the non-singular
strata of the manifold $\hat{\cal M}_I$.

Although this provides  a working definition of a
superpotential $W$ there are still two points which should be
clarified in order to complete the proof of the existence of a
superpotential reproducing the syzygies. First of all we have not
proved that the polynomials $W_{i}, i=1\dots p$, corresponding to $p$
different non-singular strata coincide on the singular sets where
the closures of the strata meet.
Assuming that this is the case allows to extend the polynomial
$W$ to the singular strata by analytic continuation.
 Secondly we have not shown that the zeros of the gradient of the polynomial $W$,
when extended to the whole set of gauge invariants from the
tubular neighbourhood of $\hat{\cal M}_I$   by analytic continuation,
coincide with $\hat{\cal M}_I$. These two points require some global
analysis which goes beyond the scope of the present paper.   

In general, one can a write a IR superpotential $W(I^a)$ which is
invariant under the global symmetries of the UV theory. The ambiguities
in its construction can be resolved by means of the decoupling
technique. Two examples are discussed in the next section.

Let us now consider the cases where the $R$ symmetry belongs to
the residual flavour symmetry. This is feasible only in the case
$\mu=\mu_{adj}$ when the $R$ charges of the scalars $z_i$ vanish.
Consequently the $R$ charges of the invariants and of the
syzygies vanish.
The $(R-1)$ charges of the syzygies and of the invariants are all
$(-1)$ implying that the normal space cannot be real. The matching
condition for the $R$ symmetry is violated.
Let us focus on one particular invariant syzygy, one can
associate to this syzygy a new invariant $X$ whose role is to
cancel the anomaly. Introducing the $X$ field corresponds to an extension of the
space of invariants. 
 This field is such that its $(R-1)$ symmetry
is unity in order  to render the normal space real.
Having paired the syzygy with the field $X$ in a real normal
space one can define the invariant polynomial \eqref{superpote} by
integration along the $X$ direction. This leads to $W=XS$. This
superpotential reproduces the classical moduli space ${\cal M}_I$
augmented with the $X$ direction at the origin. In general this
superpotential is 
deformed by quantum effects which forbids the unwanted $X$
direction.

As the superpotential that we have constructed  is a polynomial we can
deduce conditions on $\mu$.  Replacing the $I^a(z^i)$  polynomials, the
IR superpotential $W$ is a polynomial in the fields $z^i$.  Now by
comparing with the general structure (\ref{Wgeneral}) required by the
$R$-symmetry, one obtains the following condition:

{\it A necessary condition for the matching of the anomalies in theories with non-trivial syzygies is:}
\be
\mu = \mu_{adj} + k, \ \ \ \ \ k=0,1,2. \label{conf}
\ee 
This is the confinement condition\footnote{Strictly speaking this only
applies to gauge theories without a UV superpotential. In the presence
of a UV superpotential, the results stand for the unbroken gauge and
flavour symmetries and the massless states even though the anomaly
matching, as previously noticed, can be analysed at the origin of the
moduli space. Indeed the massive states decoupled below the symmetry
breaking scale are in a real representation of the residual flavour
group.} for theories with $\mu \ge \mu_{adj}.$  This condition has been
already introduced in the literature \cite{Csaki:1997zb, Csaki:1997sm,
Csaki:1998th, Skiba:1998eb, Grinstein:1998zv, Grinstein:1998bu} by
requiring a polynomial structure in (\ref{Wgeneral}).Here, it follows
from the necessary 't~Hooft matching conditions on anomalies. In
particular this excludes all higher order Dynkin index theories where
the possible superpotential would have a branch cut at the origin.

Notice that we are including in the series (\ref{conf}) the theories
with quantum modified moduli spaces \cite{Seiberg:1994bz,
Poppitz:1996fh, Grinstein:1998zv, Grinstein:1998bu}, where the (quantum
modified) syzygies are usually introduced as constraints rather than as
field equations. We have shown how the case $\mu=\mu_{adj}$ can be
included by introducing a new field $X$ when the syzygies are
invariant under the flavour group.  With
this in mind, we find it more appropriate to include these theories among
the other confining theories in (\ref{conf}).

Another important remark concerning the quantum modified moduli spaces
is the following. By writing the general superpotential in terms of the
different invariants there are ambiguities due to the presence of
different combinations of invariants with the right quantum numbers.
When the superpotential is obtained through decoupling in the descent
procedure, the constraint corresponding to the equations of motion
turns out to be the quantum modified one \cite{Seiberg:1994bz}. The
syzygies follow if one discards the dimensionful scale, which is
obviously absent in the classical syzygy. Of course, the deformation of
the manifold \MI\ when taking into account the quantum modified syzygy
must preserve the isomorphism (\ref{iso}).  However, we do not have a
general proof of this conjecture.

%%%%%%%%%%%%%%%%%%%%%%%%%%%%%%%%%%%%%%%%%%%%%%%%%%%%%%%%%%%%%%%%%%%%%%%%%%%%%%%%
\section{A Case Study:  $SU(N_c)$ and $Spin(7)$ Gauge Theories}
\label{sec:examples}

Let us illustrate these general results by a study of the well-known
case of the supersymmetric $SU(N_c)$ gauge group with $N_f$ flavours of
quarks and antiquarks, denoted by $Q_\alpha^i$ and ${\tilde Q}_\alpha
^i$, respectively.  The global symmetries are \Fsu with $Q_\alpha^i$
transforming as $(N_f,1,1,1-N_c/N_f)$ and ${\tilde Q}_\alpha ^i$ as
$(1,N_f,-1,1-N_c/N_f)$.  The chiral ring is generated by meson and
baryon composite fields (with obvious contractions of the colour
indices)~:
\be
M^{i\tj} & = & 
Q^i {\tilde Q}^{\tj} \ ,
\nonumber \\
B_{i_1\ldots i_{N_f-N_c}} & = & 
\epsilon_{i_1\ldots i_{N_f}} Q^{i_{N_f-N_c+1}} \ldots Q^{i_{N_f}} 
\ ,\label{BM}\\
{\tilde B}_{\ti_1\ldots \ti_{N_f-N_c}} & = &
\epsilon_{\ti_1\ldots \ti_{N_f}}  
{\tilde Q}^{\ti_{N_f-N_c+1} } \ldots {\tilde Q}^{\ti_{N_f}}
\ ,\nonumber
\ee
constrained by the (overdetermined) set of syzygies:
\be
B_{i_1\ldots i_{N_f-N_c}}M^{i_1 \tj}
&=&0\ ,\nonumber \\
M^{i\tj_1} {\tilde B}_{\tj_1\ldots \tj_{N_f-N_c}}
&=&0 \ ,
\label{syzSU}\\
B_{i_1\ldots i_{N_f-N_c}}
{\tilde B}_{\tj_1\ldots \tj_{N_f-N_c}}&=& 
\epsilon_{i_1\ldots i_{N_f}} \epsilon_{\tj_1\ldots \tj_{N_f}}
M^{i_{N_f-N_c+1} \tj_{N_f-N_c+1}} \ldots 
M^{i_{N_f} \tj_{N_f}} \ .
\nonumber
\ee
if $N_f \geq N_c$.  In this case one can easily solve the syzygies. The
solutions are up to a $U(N_f)\times U(N_f)$ transformation given by
\be
&&M=\hbox {diag} (M_1\ldots M_{N_c} ,0\ldots 0)\ ,
\nonumber \\
&&B_{N_c+1 \ldots N_f} = \mbox{B}\ ,
\label{solSU} \\
&&{\tilde B}_{N_c+1 \ldots N_f} =
{\tilde {\mbox{B}}} \ ,
\nonumber
\ee
with all other components of $B$ and $\tilde B$ vanishing. The syzygies
(\ref{syzSU}) are satisfied by (\ref{solSU}) provided~:
\be
\mbox{B} {\tilde {\mbox{B}}} = \prod^{N_c}_{i=1} M_i \ .
\ee
The global symmetry preserved by the solution of the syzygies
(\ref{syzSU}) is $H_{z} = SU(N_f-N_c)^2 \times U(1)^{N_c-1} $ for $N_f
\geq N_c$.  The quark representation decomposes as $N_c$ copies of the
fundamental representation of $SU(N_f-N_c)$. The tangent space to the
moduli space defined by the action of $U(N_f)\times U(N_f)$ on
(\ref{solSU}) is defined by the action of the generators
$\left(\frac{U(N_f)}{U(N_f-N_c)\times U(N_c)}\right)^2$ on $M$.  This
gives two $N_c\times (N_f-N_c)$ matrices, transforming as $N_c$ copies
of the fundamental representations of each $SU(N_f-N_c)$ factor in
$H_z$.  This proves that the $SU(N_f-N_c)$ anomalies match.  Similarly
the $U(1)$'s are vector-like and the anomalies match.  This is a
consequence of isomorphism between the moduli spaces defined for
$Q_\alpha^i$, ${\tilde Q}_\alpha^i$ and the $M$'s, $B$'s and $\tilde
B$'s that fulfill (\ref{syzSU}).

{}From the global symmetries, one deduces the superpotential
\be
W=(N_f-N_c)
{\left(
\frac{ \det M - \tr \left( B 
\parallel M^{N_f-N_c} \parallel {\tilde B} \right)}%
{\Lambda_{N_f} ^{3N_c-N_f}} \right)}^{1/{(N_f-N_c)}}
\ee
where $\parallel M^{N_f-N_c} \parallel^{i_1\ldots
i_{N_f-N_c},\tj_1\ldots \tj_{N_f-N_c}}$ is the minor of rank $N_f-N_c$
of the meson matrix $M$ specified by the appropriate rows and columns.
This superpotential  has the required charge $R=2$.  Starting from $N_f
>N_c+1$, and adding a source to the $N_f$ flavour $m M_{N_f N_f}$ one
can check that it correctly satisfies the holomorphic decoupling
condition yielding the same superpotential with one less flavour, and
$m\Lambda_{N_f}^{3N_c-N_f}=\Lambda_{N_f-1}^{3N_c-N_f-1}$.  In
particular, going down to $N_f=N_c+1$, one gets $W= (\det M-B M \tilde
B) / \Lambda_{N_c+1}^{2N_c-1}$ and decoupling the $N_c+1$ flavour the
superpotential yields the quantum constraint of the $N_f=N_c$ theory,
$W= X(\det M-B\tilde B-\Lambda_{N_c}^{2N_c})$ with the help of a
Lagrange multiplier $X={M_{N_fN_f}}/{\Lambda_{N_c+1}^{2N_c-1}}$.  For
$N_f=N_c, N_c+1$ the equations for the extrema of the superpotential
reproduce the syzygies.  When $N_f > N_c+1$, the superpotential is not
polynomial and it does not yield the syzygies.

%%Spin7
\myskip

The previous example dealt with a vector-like theory with the matter
fields in a real representation of the gauge group.  In the following
we will discuss the case of $Spin(7)$ with $N_f$ spinors in the $8$
representation \cite{Pouliot:1995zc}. In order to describe the low
energy dynamics of the theory let us study the global symmetries. There
is a non-Abelian flavour symmetry $SU(N_f)$ and also a single anomaly
free $U(1)_R$ symmetry which is non-anomalous. The spinors have charges
$(N_f,1-\frac {5}{N_f})$. The gauge invariants are the mesons
$M^{ij}=q^iq^j$ which is a symmetric matrix and the baryons
$B^{i_1i_2i_3i_4}= q^{[i_1}q^{i_2}q^{i_3}q^{i_4]}$.  It is also
convenient to define the baryons as the Hodge dual of $B$, \ie\ as an
$(N_f-4)$ antisymmetric tensor.

There are two different syzygies.  The first one is formally
$\parallel M^5 \parallel = M BB$ or explicitly
\be
\epsilon^{i_1 \ldots i_{N_f}} 
\epsilon^{j_1 \ldots j_{N_f}}
M_{i_1j_1} \ldots M_{i_5 j_5}=
M_{ij}B^{ii_6 \ldots i_{N_f}} 
B^{jj_6 \ldots j_{N_f}} .
\ee
\noindent The second one denoted by $M^2 B +B^2 =0$ reads 
\be
M_{i_1 j_1} M_{i_2 j_2} B^{i_1 i_2 j_3 \ldots j_{N_f-4}}
 + \epsilon^{a b c d \,j_3 \ldots j_{N_f-4}} B^{ef}{}_{ab} B_{efcd}
=0 .
\ee

\noindent Let us define the gauge invariant polynomial
\begin{multline}
\Omega_{N_f}=\det M + 
a_{N_f} B^{i_1 \ldots i_{N_f-4}} B^{j_1 \ldots j_{N_f-4}} 
M_{i_1 j_1} \ldots M_{i_{N_f-4} j_{N_f-4}}\\
+ b_{N_f} \epsilon^{i_1 \ldots i_{N_f}} B_{i_1 i_2 j_1 \ldots j_{N_f-6}}
B_{i_3 i_4 k_1 \ldots k_{N_f-6}} B_{i_5 \ldots i_{N_f}}
M^{j_1 k_1} \ldots M^{j_{N_f-6 } k_{N_f-6}}
\end{multline}
where $a_{n-1}=(n-4) a_n,\ b_{n-1}=(n-4)(n-6)b_n$ and the initial values
$a_6=-1,\ b_6=-1$.
There is a unique superpotential
\be
W_{N_f} = (5-N_f) 
\left( 
\frac{\Omega_{N_f}}{\Lambda_{N_f}^{15-N_f}}
\right)^{1/(N_f-5)}
.
\ee
which has a $R$-charge of two and satisfies the decoupling
requirements.  It reproduces the syzygies when $N_f=6$. This
corresponds to the anomaly matching case and $\mu=\mu_{adj}+1$ as
$\mu_{spinor}=1,\ \mu_{adj}=5$.  For $N_f$, the anomalies match when
including the Lagrange multiplier charged under $U(1)_R$.  For
$6<N_f<15$ the superpotential is not polynomial and therefore there is
no matching.

We have thus given two examples where the  possibilities
$\mu=\mu_{adj}+k,\ k=1,2$ for the index of the matter fields have been
illustrated.

%%%%%%%%%%%%%%%%%%%%%%%%%%%%%%%%%%%%%%%%%%%%%%%%%%%%%%%%%%%%%%%%%%%%%%%%%%%%%%%%
\section{Syzygies and Duality}
\label{sec:duality}

As shown in the previous section the gauge theories with $\mu >
\mu_{adj}+2$ are far more difficult to analyse than their counterparts
with either no syzygies or a superpotential describing the low energy
physics. Let us first recall the conventional lore
\cite{Intriligator:1996au, Peskin:1997qi} about gauge theories with
higher indices and their relation with syzygies.

It is generally assumed that theories with $2+\mu_{adj}<\mu<3\mu_{adj}$
have an infrared fixed point \cite{Seiberg:1995pq} where they are
described by a superconformal theory.  This fixed point describes
either  an interacting theory or it is a free theory in magnetic
phase.  Using this conjecture one can describe the set of chiral
primary fields \footnote{ Let us recall that chiral fields are defined
by $\bar D \Phi=0$ where $\bar D$ is the supersymmetric fermionic
derivative. Primary fields are such that $\Phi\ne \bar D \chi$ for a
given superfield $\chi$.} of the superconformal theories.  The primary
chiral superfields form a chiral ring under the operator product
defined by the short distance expansion. This ring is identified with
the ring of gauge invariant polynomials. The syzygies are therefore
exact quantum relations between the chiral primary fields at the
superconformal fixed point.

At these fixed points, the dimension $d$ of the chiral primary field is
related to the $R$-charge by $d=\frac {3}{2} |R|$.  Unitary conditions
restrict these dimensions to be greater than one, the bound being
saturated for a free field.  So the description in terms of infrared
fixed points breaks down when the dimensions of certain operators
becomes formally less than one. For instance for vector-like theories
with quarks and antiquarks this happens for the mesons when $\mu<\frac
{3}{2}\mu_{adj}$.  In that case the theory is supposed to possess a
dual description where the dual mesons become a free field, leading to
a ``free magnetic phase". Unfortunately there is no prescription for
constructing such dual models.  Nevertheless  the known examples have
to satisfy  stringent consistency checks.

In this section, we present, for two well known examples, a new check.
Whereas in ``electric theories" the  gauge invariants are composite
fields and so subject to  syzygies, in the magnetic dual theories, some
of them appear as elementary fields and do not have \apriori\ to
satisfy the syzygies.  As suggested in \cite{Intriligator:1996au,
Aharony:1995qs}, these ``magnetic syzygies" should appear as
non-perturbative effects, which we will show explicitly by an
appropriate regularization of the superpotentials generated non
perturbatively in the magnetic theories.

Let us first describe the duality of the $SU(N_c)$ with $N_f>N_c+2$
quarks and antiquarks. The dual gauge group \cite{Seiberg:1995pq} is
$SU(N_f-N_c)$ with $N_f$ matter fields $q,\tilde q$ in the fundamental
and antifundamental representations respectively. The dual theory also
possesses dual mesons $M_D$, invariant under $SU(N_f-N_c)$ and that
couple to $q$ and $\tilde q$ through a superpotential:
\be
W= {\frac{1}{\mu}} \ \tr \left( M_D q {\tilde q}^t \right),
\ee
where $\mu$ is a scale parameter.  This superpotential is not
sufficient to show the equivalence between the original theory and its
dual, and in particular is not sufficient to identify the dual mesons
$M_D$ with the electric ones $Q{\tilde Q}$.  This requires non
perturbative modification of the magnetic moduli space.  Along a flat
direction with $\langle M_D \rangle \ne 0$, the matter fields $q$ and
$\tilde q$ become massive and decouple from the low energy theory.  The
pure gauge theory then undergoes gaugino condensation which generates
an extra contribution to the superpotential \cite{Aharony:1995qs}
\be
 W_{n.p.}=(N_c-N_f)
\left( \frac {\det M_D}{\Lambda^{3N_c-N_f}} \right)^{1/(N_f-N_c)}
.
\label{Wnpert}
\ee
The analysis of the vacua of the dual theory necessitates to introduce
source terms $W_{reg.} = \tr  (mM_D)$ in the superpotential. The vacua are
obtained by taking the limit of zero sources.  {}From the total
superpotential $W+W_{n.p.}+W_{reg.}$, the equations of motion 
for the mesons lead to a vacuum satisfying
\be
M_D=\left( \Lambda^{3N_c-N_f} 
\det \left(m+{\frac {q{\tilde q}^t}{\mu}}\right)
\right)^{1/N_c}
 \left( m+{\frac {q{\tilde q}^t}{\mu}}\right)^{-1}
.
\label{vacMD}
\ee
As in the electric theory, a flavour and gauge transformation
diagonalize the matter fields
\be
(q^i_\alpha, {\tilde q}^i_\alpha) =
\begin{cases}
(a_i, {\tilde a}_i) \delta_{\alpha}^i & \text{for $i \leq N_f-N_c$},\\
0 & \text{for $i > N_f-N_c$}.
\end{cases}
\ee
The matrix parameter $m$ regularizes the matrice on the right hand
side of \eqref{vacMD}. In the limit $m \to 0$, the vacuum (\ref{vacMD}) becomes 
\be
M_D=
\left( \Lambda^{3N_c-N_f} \prod_{i=1}^{N_f-N_c} 
{\frac{a_i {\tilde a}_i}{\mu}}
\right)^{1/N_c}
\left(
\ba{cc}
0&0\\
0& m_D
\ea
\right)
,
\label{MD}
\ee
where $m_D$ is a $N_c \times N_c$ matrix of determinant one.  The claim
is that this relation exactly reproduces the syzygies of the electric
theory. Indeed, according to ref.\cite{Intriligator:1996au}, the
identification between the electric and magnetic gauge invariants is
\begin{equation}
\begin{split}
M &\to  M_D,\\
(B,{\tilde B})^{i_1\ldots i_{N_f-N_c}} 
&\to  
\sqrt {\Lambda^{3N_c-N_f}\mu ^{N_c-N_f}}\ 
(b, {\tilde b})^{i_1\ldots i_{N_f-N_c}},
\end{split}
\label {dual}
\end{equation}
where $b^{i_1\ldots i_{N_f-N_c}}$ is defined by the totally
antisymmetric gauge invariant combination $q^{[i_1} \ldots
q^{i_{N_f-N_c}]}$.  Using  the electric variables, the relation
(\ref{MD}) reads
\be
B{\tilde B} = \parallel M^{N_c} \parallel,
\ee
whereas the two other sets of electric syzygies are also trivially
satisfied.  In that sense, the magnetic theory with an appropriate
regularization  satisfies the same set of syzygies as the electric
theory.

%%%%
\vspace{.5cm}
The same analysis can be applied to the $Spin(7)$ theory. Its dual
theory \cite{Pouliot:1995zc} is a chiral $SU(N_f-4)$ gauge theory with
$N_f$ matter fields $q^i$ in the antifundamental representation and one
field $S$ in the symmetric representation. There is also a symmetric
meson matrix $M_D$ which is a singlet under the gauge group.  These
fields are coupled by the superpotential
\be
W=\frac{1} {\mu^2} \tr (M_D q S q^t)+
{\frac{1}{\mu^{N_f-7}}} \det S
.
\ee
This tree-level superpotential needs to be completed by a
non-perturbative part and source terms.  As before, we consider a flat
direction where the symmetric field is proportional to the identity and
$M_D\ne 0$. The matter fields become massive leading to gaugino
condensation of the pure gauge theory. This leads to the following term
in the superpotential
\begin{equation}
\begin{split}
W_{n.p.}
& 
= (5-N_f)
\left( \frac {\det M_D}{\Lambda^{15-N_f}} \right)^{1/(N_f-5)}
\\
W_{reg.}
& 
= 
\tr (m M_D) + \tr  (m_s S)
,
\end{split}
\end{equation}
where sources $m$ and $m_s$ have been introduced for regularization
purposes.  Firstly,  the $D$-terms have to vanish, \ie\,
\be
2S^{\dagger}S-q^{\dagger}q= \lambda\, {\mathbf{1}} ,
\ee
with a suitable real number $\lambda$.  Then once again, gauge and
flavour transformations can simultaneously diagonalize $q$ and $S$.
The solutions of equations of motion derived from the total
superpotential are,
\be
	\label{vacuaS}
S = 
\left( -\mu^{N_f-7}\,
\det \left( m_s +\frac{M_D qq^t}{\mu^2} \right) \right)^{1/(N_f-5)}
\ \left(m_s + \frac{M_D qq^t}{\mu^2} \right)^{-1},
\ee
and
\be
	\label{vacuaMD}
M_D = 
\left( \Lambda^{15-N_f}
\det \left( m +\frac{qSq^t}{\mu^2} \right) \right)^{1/5}
\ \left(m + \frac{qSq^t}{\mu^2} \right)^{-1}
.
\ee
The physical vacua correspond to the limit $m \to 0$ and $m_s \to 0$
where \eqref{vacuaS} and \eqref{vacuaMD} become
\be
S = \left( -\mu^{N_f-7}\,
\frac{M_{N_f-4}\, q_{N_f-4}^2}{\mu^2} \right)^{1/(N_f-5)}
\left(
\ba{cc}
s & 0 \\
0&0
\ea
\right)
,
\label{sSpin7}
\ee
$s$ being a $(N_f-5)\times(N_f-5)$ matrix of determinant one,
and 
\be
M_D =  
\left( \Lambda^{15-N_f}
\prod_{i=1}^{N_f-5} \frac{q_i^2 S_i}{\mu^2} \right)^{1/5}
\left(
\ba{cc}
0 & 0\\
0 & m_D
\ea
\right)
,
\label{MDSpin7}
\ee
$m_D$ being a $5\times5$ matrix of determinant one.  Quarks and mesons
have only one non-vanishing flavour index in common that is taken here
to be the $(N_f-4)^{\mbox{{\small th}}}$ one.  Equations (\ref{sSpin7})
and (\ref{MDSpin7}) assures that only one $5\times5$ minor of $M_D$ is
non vanishing and it is given by
\be
\parallel M_D^5 \parallel = -\Lambda^{15-N_f} \mu^{1-N_f} M_{N_f-4}\ bb,
\label{sysmagn}
\ee
where $b$ is the only non vanishing dual baryon $b=\prod_{i=1}^{N_f-4}
q_i$.  Providing the following identification between electric and
magnetic gauge invariants
\begin{equation}
\begin{split}
M & \to  M_d,\\
B^{i_1\ldots i_{N_f-4}} &\to  \sqrt{- \Lambda^{15-N_f} \mu^{1-N_f}} 
\ b^{i_1\ldots i_{N_f-4}}, 
\end{split}
\end{equation}
the relation (\ref{sysmagn}) is the only non trivial electric syzygy
expressed in terms of magnetic variables.\\

So we have seen in the two examples explicitly studied how 
the syzygies of the electric theory are fullfilled in the magnetic theory 
thanks to non-perturbative effects that eliminate magnetic vacua without
counterparts in the electric theory.

%%%%%%%%%%%%%%%%%%%%%%%%%%%%%%%%%%%%%%%%%%%%%%%%%%%%%%%%%%%%%%%%%%
\section{Geometry of the Moduli Spaces} 
\label{sec:geometry}

In supersymmetric theories, the K\"ahler potential is sensitive to
perturbative quantum effects, non-perturbative ones, threshold and
decoupling corrections. For this reason it is not obvious how to extend
the analysis of confinement and duality to this sector of the theory.
Nevertheless, the K\"ahler potential encodes important information on
the dynamics.  It gives the $\sigma$-model geometry of the complex
scalar manifolds, with the quantum modifications that includes a scale
dependence.  Therefore, one would like to understand the relationship
between the K\"ahler structures of dual pairs of theories, for
instance. Of course, a classical approach would be incomplete, but it
could be of some usefulness if only the perturbative corrections are
the most  relevant, which is consistent with the fact that we do not
expect quantum modifications of the moduli spaces in dual theories.

In this section, we present an approach to the geometry of the moduli
space \cite{Brax:1997cf} which is based on a study of the hidden
symmetries of the scalar potential. This provides a parameterisation in
terms of IR degrees of freedom. We show that the moduli space consists
of one orbit of a non-compact group ${\cal F}$ (the so-called {\it
commutant} as it commutes with the gauge symmetry) and its closure,
corresponding to the singular orbits. The commutant group is a
non-trivial non-compact extension of the compact flavour symmetry.
Therefore the geometry of the moduli space is that of a K\"ahler coset
space ${\cal F}/{\cal H},$ where ${\cal H}$ is the little group of the
orbit.

Of course this structure is to be identified in both the dual theories,
providing IR parameterisations along the lines of the usual approach to
spontaneously broken symmetries.  This would give access to a
comparison of the classical K\"ahler potential\footnote{For different
attempts see refs.\cite{Poppitz:1994tx, Cho:1998vc}.} of the dual
theories, but there is a major obstacle in the fact that the
parameterisations are not holomorphic. In spite of this failure, the
possibility of identifying (in both senses) the geometry of the moduli
space of dual theories is worthwhile presenting.

We shall carry out our analysis in the explicit context of the
$SU(N_c)$  gauge theory with $N_f$ flavours of quarks and antiquarks.
The classical moduli space is not modified non-perturbatively when
$N_f>N_c$.  The construction of the duality between the electric and
the magnetic theories  has been obtained in the conformal range
${\frac{3}{2}}N_c<N_f<3 N_c$ in which the electric and magnetic
theories are both asymptotically free.  We denote by $Q^i_\alpha$ and
${\tilde Q}^{\ti }_{\ta}$ the corresponding scalar fields.  The
K\"ahler potential is
\be
{\cal K} (Q, Q^\dagger, {\tilde Q} , {\tilde Q}^\dagger )
=\tr ( Q^\dagger Q + {\tilde Q}^\dagger {\tilde Q}) \ .
\label{K_electric} 
\ee

In order to identify this manifold, one has to determine the symmetries
of the flat potential condition $V=0$, namely the simultaneous zeros of
the $D^A$'s and $F_i$'s. In the electric theory, due to the absence of
superpotential, the vacua are only restricted to the $D$-flatness
equations which are equivalent to the single relation:
\be
(Q^i_\alpha)^* Q^i_\beta 
- {\tilde Q}^{\ti }_{\ta}
({\tilde Q}^{\ti }_{\tilde \beta})^*
=  \lambda \, \delta_{\alpha\beta} \ ,
\label{Q=1}
\ee 
\noindent where $\lambda$ is  a real  number.  Equation (\ref{Q=1})
explicitly exhibits an $SU(N_f,N_f)\times U(1)_B$ flavour invariance
\cite{Buccella:1982nx}, acting on the $2N_f$ component vectors $(Q\, ,
\, {\tilde Q}^* )$, including a Cartan generator corresponding to the
$U(1)_R$. Moreover there is an obvious invariance under dilation.
Finally the symmetry of the moduli space  is
\be
{\cal F}_e = U(N_f,N_f) \times D  \ .
\ee
\noindent Notice that holomorphy is not preserved by the action of
${\cal F}_e$.  This is the main drawback of this approach where the
supersymmetry is not explicit.

The moduli space corresponds to the following  orbits of ${\cal F}_e$.
There are two conjugated {\it baryonic\ orbits} corresponding to
$\lambda>0$ and $\lambda<0$. They are called baryonic orbits as there
is always a non-zero baryon on these branches of the moduli space.
Each of them forms a single orbit under  the symmetry group ${\cal
F}_e$.  The case $\lambda<0$ is obtained from $\lambda >0$ by
exchanging the roles of $Q$ and  ${\tilde Q}^*$. Their common boundary
$\lambda=0$ is the {\it mesonic orbit}.

Therefore, the vacua manifold has been identified as a finite set of \\
$SU(N_f,N_f)\times U(1)_B \times SU(N_c)\times D$ orbits.  Each of
these orbits can be represented by the quotient of the symmetry group
(acting transitively on the orbit) by the stabilizer (or little group)
of one point. Thus we have to identify the stabilizer of each
representative point considered earlier to characterize each orbit.

For the {\it baryonic orbit}, the little group associated to the
generic vacuum takes a simple structure of direct product
\be
{\cal H}_e = SU(N_f-N_c,N_f)\times U(1) \times SU(N_c)_D \ ,
\ee
where $SU(N_c)_D$ is the diagonal  combination of the gauge $SU(N_c)_G$
and an $SU(N_c)$ subgroup of $U(N_f,N_f)$. The $U(1)$ is a combination
of the $U(1)_B$ and an element of the  Cartan subalgebra of
$SU(N_f,N_f)$.  Here the gauge group is completely broken; then after
eliminating  the spurious massless scalars associated to the Higgs
mechanism, the real dimension of the coset ${\cal F}_e/{\cal H}_e$ is
$4N_fN_c - 2N_c^2 +2 $.

In order to further characterize the geometry of the baryonic branch of
moduli space, we first extract a flat subspace associated  to the
diagonal $Gl(1,\C)$ factor in the coset and then introduce as
coordinates the $N_C\times (2N_f-N_c)$ complex  matrix $z$ and define
its transformation under $U(N_f,N_f)$ as
\be
z \to (Az+B) (Cz+D)^{-1} \ .
\label{homographic}
\ee
The elements of the  coset are then parameterized by the exponentials 
$e^{t(z)}$
\be
\left(
\begin{array}{c}
Q\\
{\tilde Q}^*
\end{array}
\right)_{\vert D^A=0}
\, = \, 
e^{t(z)}
\left(
\begin{array}{c}
{\mathbf {1}}_{N_c}\\
0
\end{array}
\right)\ ,
\label{parameter}
\ee
 where
\be
t(z)=
\left(
\begin{array}{cc}
0 & z \\
-\eta z^\dagger & 0 
\end{array}
\right)
\ .
\label{exponent}
\ee  

\noindent and  $\eta$ is the $(N_f-N_c,N_f)$ signature.  There is only
one $U(N_f,N_f)$-invariant K\"ahler potential  up to a  K\"ahler
transformation, namely: ${\cal K} = \tr \ln (1+z \eta z^\dagger)$ .
Nevertheless as $U(N_f,N_f)$ is not a symmetry of the theory (it does
not preserve the kinetic terms), this is not the K\"ahler potential of
the theory on the moduli space.

Let us now check the correspondence between the massless scalar fields
and the moduli fields. As already discussed previously, since there is
no superpotential and the whole gauge group is Higgsed around the
vacuum, all the scalars are moduli with the exception of the $N_c^2-1$
complex scalars given by $T^A_{i\beta}  Q^i_\beta $, associated to the
$SU(N_c)$ massive vector multiplets.  All the $4N_fN_c-2N_c^2+2$
remaining scalar fields are massless, and their number coincides with
the real dimension of $G_e/H_e$.

{}For the {\it mesonic orbit}, the pattern is more complicated since
the little group associated to the generic vacuum now has a structure
of semi-direct product, ${\cal H}_r\times SU(N_c-r)_G$, with:
\be
{\cal H}_r =  SU(N_f-r,N_f-r)\times U(1)^2\times SU(r)_D
\rtimes {\tilde H} \ ,
\label{stab_elec_meson}
\ee
where $SU(r)_D$ is the diagonal  combination of the  gauge and flavour
$SU(r)$ subgroups,  the two $U(1)$'s are combinations of the $U(1)_B$,
an element of the  Cartan subalgebra of $SU(N_f,N_f)$ and an element of
the Cartan subalgebra of $SU(N_c)_G$. The semi-direct factor ${\tilde
H}$ is a nilpotent subgroup~\footnote{ Just as for the Lorentz group,
the little groups of singular orbits of non compact groups have semi
direct products with nilpotent subgroups.} with generators transforming
as $(2(N_f-r), \overline{r}) \oplus (\overline{2(N_f-r)},r)$ under
$U(N_f-r,N_f-r)\times SU(r)_D$ and the Abelian subalgebra defined by
their commutators, which transform as $(1, Adj\oplus 1)$ .

The mesonic orbits are stratified by the index $r$.  The stratum
corresponding to  $r=N_c$ is such that the gauge group is completely
broken.  In fact the mesonic orbits correspond to the ``infinitely
boosted'' baryonic orbit.   Hence the moduli space is the closure of
the baryonic orbit,  \ie\ by applying appropriate boosts and a global
dilation to  the baryonic orbit one can converge to  all the strata of
the mesonic orbits.  In the same way, by applying infinite boosts, we
can go from a mesonic orbit with a stratification index $r$ to a more
singular one with lower index.  Geometrically, these boosts correspond
to the shrinking of some circles in the moduli space.  {}From a
physical point of view, the stratification index is  related to the
number of massless singlets of the theory.  As the stratification index
goes  from $r$ to $r-1$, the corresponding orbits differ by a
dimension  $4(r-N_f)-2$.  In terms of the solutions of the syzygies,
the baryonic orbit is equivalent to the open set characterized by $b\ne
\tilde b$ while the mesonic orbits correspond to $b=\tilde b$. One
retrieves the fact that the mesonic orbits form the natural boundary of
the baryonic orbit.

We now turn to the analysis of the dual theory, or magnetic theory,
described in the previous section. Notice in particular the
non-perturbative superpotential (\ref{Wnpert}), which restricts the
solutions so that the meson field matrix has rank less than $N_c$. This
is crucial   in identifying  the dual moduli spaces. The magnetic
theory possesses the same anomaly-free global symmetries as the
electric one $ SU(N_f) \times SU(N_f) \times U(1)_B \times U(1)_R \ ,$
which transform the dual quark superfields as $({\bar
N_f},1,N_c/(N_f-N_c),N_c/N_f)$, the antiquark ones as $(1,N_f, -N_c /
(N_f-N_c),$ $N_c/N_f)$  and the gauge-singlet superfields $M$ as
$(N_f,{\bar N_f}, 0 , 2(N_f-N_c)/N_f)$.

The classical moduli space of the magnetic theory is identified with
the solutions of the $F$- and $D$-terms.  A thorough analysis of these
equations and their relationship with the electric moduli space has
been given in \cite{Brax:1997cf}. In summary
\be
q&=&e^{t(u)}q_0\ ;\nonumber\\
\tilde q&=&0 \ ;  \\
M&=& (0\ M_0)e^{t(u)^{\dagger}}\nonumber 
\ee
where
\be 
t(u) =
\left(
\begin{array} {cc}
0&u\\
-u^{\dagger}&0\\
\end{array}
\right) \ ,
\nonumber
\ee
with $u$ a $(N_f-N_c)\times N_c$ matrix which parameterises the
coset $\frac{U(N_f)}{U(N_f-N_c)\times U(N_c)}$.  The following 
$N_c \times (2N_f-N_c)$ complex matrix can be constructed:
\be 
z_m=(u^T\ M_0^T)\ .
\label{z_m}
\ee
Consider the homographic action of the  group $U(N_f,N_f)$ on $z_m$,
analogous to  (\ref{homographic}).  The magnetic moduli space is
invariant under this non-linear action  of $U(N_f,N_f)$.  Consider the
image of the origin $z_m=(0,0)$. As the action on $z_m$ is the same as
(\ref{homographic}) on the electric baryonic orbit we know that the
image of the origin is the whole  baryonic orbit.  Therefore the
baryonic branches  of the electric and magnetic moduli spaces are
isomorphic as non-compact complex manifolds.  The dualising map reads
simply $ z \leftrightarrow z_m $.  This isomorphism is valid at the
level of the classical moduli spaces.  The non-perturbative part of the
superpotential is necessary to study another branch of the dual
theory.  In the case $q=\tilde q=0$ the mesons $M$  is  restricted by
the superpotential (\ref{Wnpert}) to have rank $N_c$.

The K\"ahler potential of the magnetic theopry is not known. If the
quark kinetic term is expected to become canonical in the UV region,
there is no reason for the  K\"ahler geometry of the meson fields to be
trivial. One can try to use the isomorphism between the baryonic orbits
of the dual theories in order to deduce the K\"ahler potential of the
dual meson field in the UV region.  However, we cannot use this twin
parameterisation of the electric and magnetic fields in the classical
part of the moduli space to deduce the magnetic K\"ahler potential from
the electric one (\ref{K_electric}).  Indeed, the induced metric cannot
be straightforwardly calculated from (\ref{parameter}) and
(\ref{exponent}) because this is not an analytic transformation of
variables consistent with the K\"ahlerian geometry.  Still, it provides
a link that we hope to be able to exploit in the future.

%%%%%%%%%%%%%%%%%%%%%%%%%%%%%%%%%%%%%%%%%%%%%%%%%%%%%
\section{Conclusions}

We have argued that the 't~Hooft matching of the anomalies, when the
normal space to the manifold defined by the syzygies is a real
representation of the flavour unbroken subgroup, is equivalent to the
existence  of a polynomial superpotential that has been put forward to
characterize the confinement in $N=1$ supersymmetric gauge theories.
This allows  a complete classification of the confining theories for
$\mu \ge \mu_{adj}$. Non-confining theories, with $3\mu_{adj}\ > \mu >
\mu_{adj}+2$, are expected to have dual(s).  Some of the properties
concerning both confinement and duality discussed in this paper have
been illustrated by the analysis of two series of theories, $SU(N_c)$
and $Spin(7)$, with a descent relation between the successive
decoupling of flavours.

For all confining theories the IR superpotentials for the gauge
invariant composites are completely fixed by the flavour symmetries. We
have shown, in theories characterized by non-trivial syzygies, 
that the following additional requirements are strongly
related:  {\it (i)} the matching of the anomalies, {\it (ii)}
polynomiality of the superpotential, and {\it (iii)} the equations of
motion leading to the syzygies.  We have stressed the equivalence of {\it
(i)} and {\it {(ii)}} by constructing a superpotential in the case of a
normal space in a real representation of the residual flavour group,
and we have partially proved the equivalence with {\it (iii)}
(we were unable to prove that the syzygies are the only solutions 
of the equations of motion). 
Of course, one would like to have also a proof of the conjecture that the
normal space to the zeros of the syzygies are not in an anomaly free
complex representation of the unbroken flavour subgroup.

In the case of dual theories, we have checked, with the aid of  explicit
examples, the consistency between duality and syzygies. This property
is verified only if  non-perturbative superpotentials are included.

Finally we have discussed how a non-compact hidden symmetry of the
vacua characterizes the classical moduli space geometry, but the
non-analyticity of the action of the non-compact hidden group makes it
difficult to gain  further insight into the K\"ahler geometry of the
dual theories.

%%%%%%%%%%%%%%%%%%%%%%%%%%%%%%%%%%%%%%%%%%%%%%%%%%%%%
\section*{Acknowledgements}

We thank the referee for constructive remarks
that contributed to the improvement of section 4.

%%%%%%%%%%%%%%%%%%%%%%%%%%%%%%%%%%%%%%%%%%%%%%%%%%%%%%%%%%%%%%%%%%%%%%%%%%%%%%
%%%%%%%%%%%%%%%%%%%%%%%%bibliographie%%%%%%%%%%%%%%%%%%%%%%%%%%%%%%%%%%%%%%%%%%
%\bibliographystyle{h-elsevier}

%\bibliography{invariant}

\end{document}